\documentstyle[psfig,nicV]{article}

\begin{document}
%
%
\heading{%
Merging Neutron Stars and Black Holes
as Sources of Gamma-Ray Bursts and Heavy Elements\\
%
}
\par\medskip\noindent
%
\author{%
H.-Thomas Janka$^{1}$, Maximilian Ruffert$^{2,3}$, Thomas Eberl$^{1}$
}
\address{
Max-Planck-Institut f\"ur Astrophysik, Postfach 1523,
D--85740 Garching, Germany
}
\address{
Institute of Astronomy, Madingley Road, Cambridge~CB3~0HA, U.K.
}
\address{
Dept.~of Mathematics \& Statistics, University of Edinburgh,
Scotland, EH9 3JZ, U.K.
}
\begin{abstract}
Hydrodynamic simulations were performed of the dynamical phase of
the merging of binary neutron stars (NS-NS) and of neutron star 
black hole binaries (NS-BH), using a physical nuclear equation of
state~\cite{janka.ls} and taking into account the emission of 
gravitational waves and neutrinos.
\end{abstract}

\section*{Simulations and some results}
NS-NS and NS-BH
mergers have been suggested as sources of $\gamma$-ray bursts 
at cosmological distances (e.g., \cite{janka.nar})
and of r-process nuclei~\cite{janka.eich}. These merger
events are estimated to occur at rates of the order $10^{-5}$ 
per year per galaxy (e.g., \cite{janka.fry}).

By means of three-dimensional hydrodynamic simulations with a Eulerian
PPM code we have investigated the dynamical phases of the 
coalescence~\cite{janka.ebe,janka.ruf}. The simulations were
performed with four nested cartesian grids which allow for both a good
resolution near the central black hole and a large computational volume.
In a post-processing step, we evaluated our models for the energy
deposition by $\nu\bar\nu$ annihilation around the merging objects.

\begin{center}
\tabcolsep=1.3mm
\begin{tabular}{l c c c c c c c c c c c c}
\multicolumn{13}{l}{{\bf Table 1.} Results of merger models.
Energies are given in units of 1foe = $10^{51}\,{\rm erg}$.}\\
\hline
\\[-3.5mm]
\multicolumn{1}{c}{model}&
\multicolumn{1}{c}{masses}&
\multicolumn{1}{c}{spin}&
\multicolumn{1}{c}{$\Delta M_{\rm d}$}&
\multicolumn{1}{c}{$t_{\rm sim}$}&
\multicolumn{1}{c}{$\left\langle L_{\nu}\right\rangle$}&
\multicolumn{1}{c}{$a_{\rm BH}$}&
\multicolumn{1}{c}{$q_{\nu}$}&
\multicolumn{1}{c}{$q_{\nu\bar\nu}$}&
\multicolumn{1}{c}{$\dot E_{\nu\bar\nu}$}&
\multicolumn{1}{c}{$t_{\rm acc}$}&
\multicolumn{1}{c}{$E_{\nu}$}&
\multicolumn{1}{c}{$E_{\nu\bar\nu}$}
\\
\multicolumn{1}{c}{ }&
\multicolumn{1}{c}{{\scriptsize($M_{\odot}$)}}&
\multicolumn{1}{c}{ }&
\multicolumn{1}{c}{({\scriptsize${M_{\odot}\over 100}$})}&
\multicolumn{1}{c}{{\scriptsize$({\rm ms})$}}&
\multicolumn{1}{c}{({\scriptsize$100{{\rm foe}\over{\rm s}}$})}&
\multicolumn{1}{c}{ }&
\multicolumn{1}{c}{\%}&
\multicolumn{1}{c}{\%}&
\multicolumn{1}{c}{({\scriptsize${{\rm foe}\over{\rm s}}$})}&
\multicolumn{1}{c}{{\scriptsize$({\rm ms})$}}&
\multicolumn{1}{c}{$\!\!${\scriptsize$({\rm foe})$}$\!\!$}&
\multicolumn{1}{c}{$\!\!${\scriptsize$({\rm foe})$}$\!\!$}
\\
\hline
NS-NS & 1.2--1.2  & solid & 2.0  & 10 & 1.5 & ---  & --- & 1.0 & 1.4  & --- & --- & ---  \\
NS-NS & 1.2--1.8  & solid & 3.8  & 13 & 2.0 & ---  & --- & 1.3 & 2.6  & --- & --- & ---  \\
NS-NS & 1.6--1.6  & anti  & 0.01 & 10 & 4.0 & ---  & --- & 1.5 & 6.0  & --- & --- & ---  \\
NS-NS & 1.6--1.6  & none  & 0.23 & 10 & 5.0 & ---  & --- & 1.8 & 9.1  & --- & --- & ---  \\
NS-NS & 1.6--1.6  & solid & 2.4  & 10 & 3.3 & ---  & --- & 2.1 & 7.0  & --- & --- & ---  \\
BH-AD & 3.0--0.26 & solid & ---  & 15 & 1.2 & 0.42 & 1.3 & 0.4 & 0.49 & 68  & 8   & 0.03 \\
BH-NS & 2.5--1.6  & solid & 0.2  & 10 & 8.0 & 0.76 & 3.0 & 2.8 & 22.4 & 40  & 32  & 0.9  \\
BH-NS & 5.0--1.6  & solid & 5.6  & 15 & 5.0 & 0.33 & 2.8 & 1.8 & 9.0  & 56  & 28  & 0.5  \\
BH-NS & 10.0--1.6 & solid & 10.0 & 10 & 2.5 & 0.14 & 2.8 & 0.9 & 2.3  & 82  & 21  & 0.2  \\
\hline
\end{tabular}
\end{center}

Results for some simulations are listed in Table~1. The NS and accretion
disk (AD) masses are baryonic masses, the spin parameter indicates
whether no spin was assumed for the NS's (``none''), or the NS spin was taken 
parallel (``solid'') or anti-parallel (``anti'') to the orbital angular 
momentum. $\Delta M_{\rm d}$ is the dynamically ejected mass, $t_{\rm sim}$
the time at the end of the simulation, 
$\left\langle L_{\nu}\right\rangle$ the typical total neutrino
luminosity, $a_{\rm BH} = Jc/(GM_{\rm BH}^2)$ the relativistic rotation 
parameter of the black hole, $q_{\nu} = L_{\nu}/(\dot Mc^2)$ the 
conversion efficiency of rest-mass energy of the gas swallowed
by the black hole into $\nu$ energy, 
$q_{\nu\bar\nu} = E_{\nu\bar\nu}/E_{\nu}$ the conversion efficiency of
$\nu$ energy into $e^\pm$ pairs, $\dot E_{\nu\bar\nu}$ the integral rate of 
energy deposition by $\nu\bar\nu$ annihilation,
$t_{\rm acc}$ the estimated duration of 
the accretion by the BH, $E_{\nu}$ the energy released in neutrinos during
$t_{\rm acc}$, and $E_{\nu\bar\nu}$ the corresponding energy converted into
$e^\pm$ pairs.

\begin{figure}[tbp]
\centerline
{\vbox{
\psfig{figure=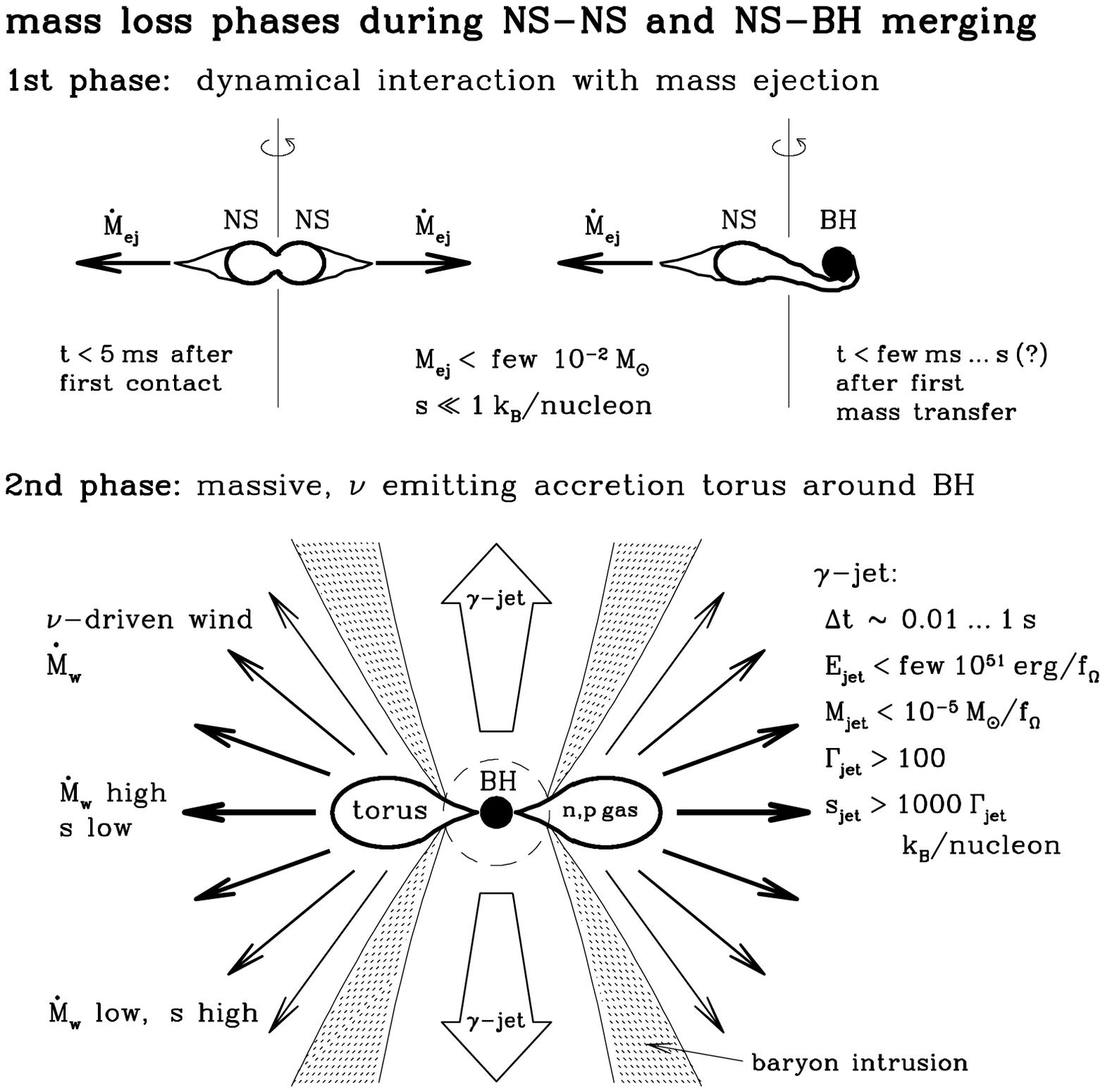,width=0.99\textwidth}}}
\caption[]{\small
Mass-loss phases during NS-NS and NS-BH merging and during the
subsequent evolution of the massive accretion disk that forms 
around the BH. 
}
\label{janka.fig1}
\end{figure}

We find that during the dynamical phase of the merging between about
$10^{-4}\,M_{\odot}$ and $10^{-2}\,M_{\odot}$ of cool gas can be
ejected (Fig.~\ref{janka.fig1}). 
The exact number depends very sensitively on the assumed
neutron star spin(s) and on the masses of the merging binary components
(Table~1). This is of potential interest for heavy-element 
nucleosynthesis~\cite{janka.ross}.

Figure~\ref{janka.fig1} sketches the result that during the dynamical 
interaction of the binary
components (1st phase) low-entropy (dense and cold) matter is
ejected from the ends of long spiral arms which are formed in the
orbital plane. During the merging the matter is heated by shocks and
viscous dissipation, and intense neutrino fluxes are emitted before
the gas is accreted into the central black hole. Energy transfer by
these neutrinos drives a high-entropy gas outflow (``neutrino wind'')
off the surface of the accretion disk. At the
same time, energy deposition by $\nu\bar\nu$ annihilation in the
essentially baryon-free funnel around the rotation axis powers
relativistically expanding $e^\pm\gamma$ jets which can give rise
to $\gamma$-ray bursts.

\begin{figure}[tbp]
\centerline
{\vbox{
\psfig{figure=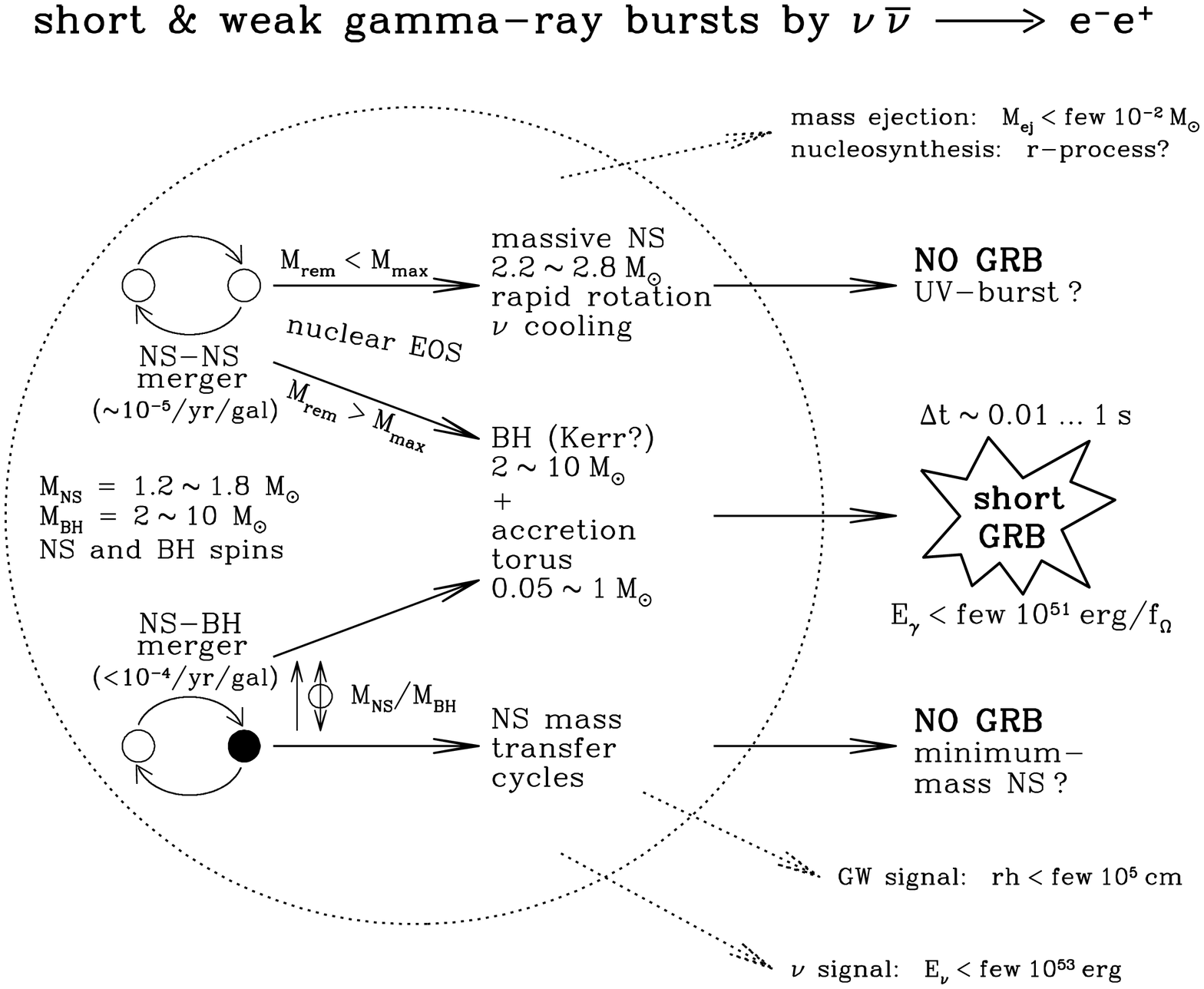,width=0.99\textwidth}}}
\caption[]{\small
Possible evolution paths from NS-NS and NS-BH mergers to short
$\gamma$-ray bursts powered by $\nu\bar\nu$ annihilation into
$e^\pm$ pairs. 
}
\label{janka.fig2}
\end{figure}

\begin{figure}[tbp] 
\centerline{
\vbox{
\psfig{figure=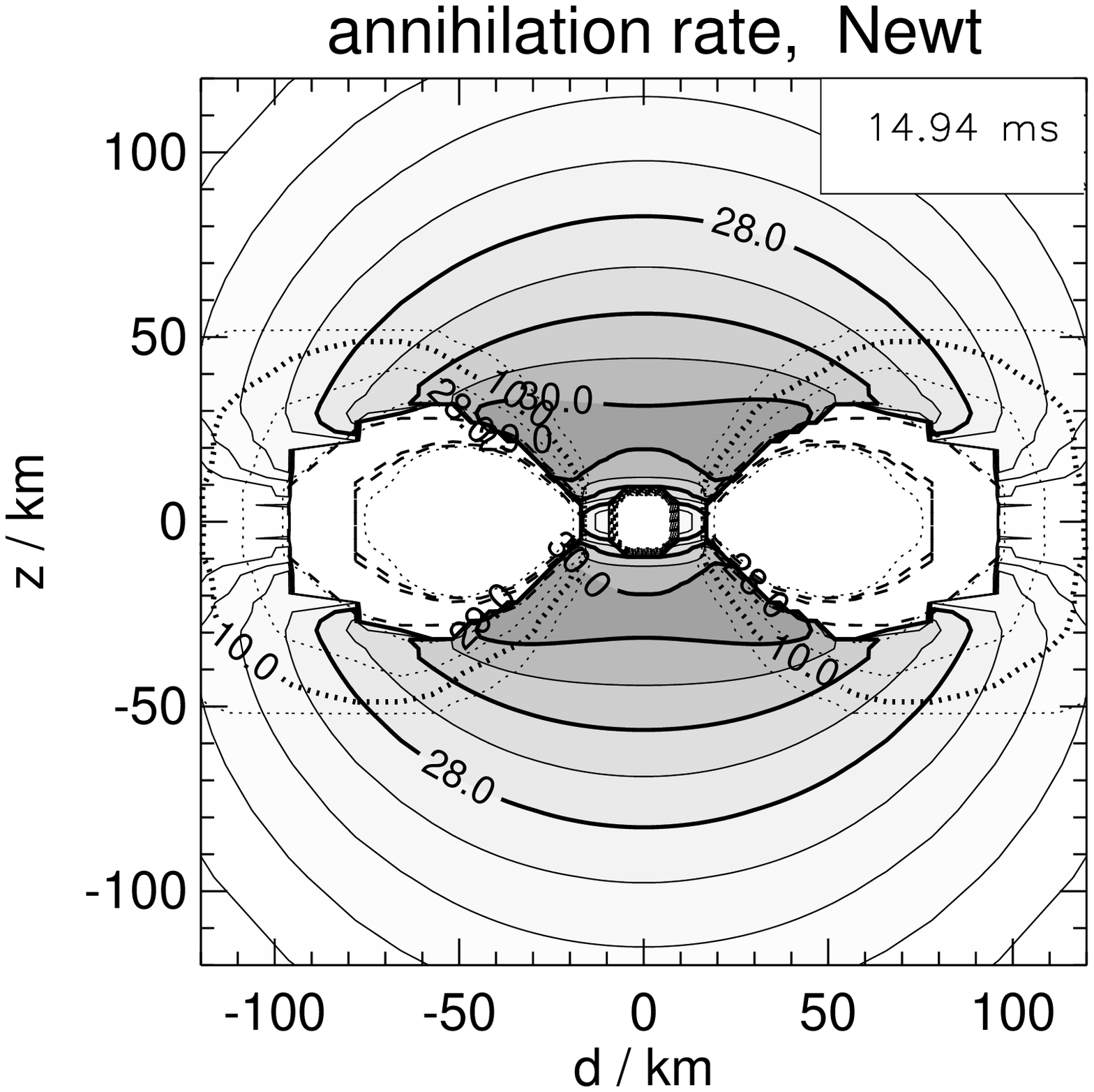,width=0.49\textwidth}
\psfig{figure=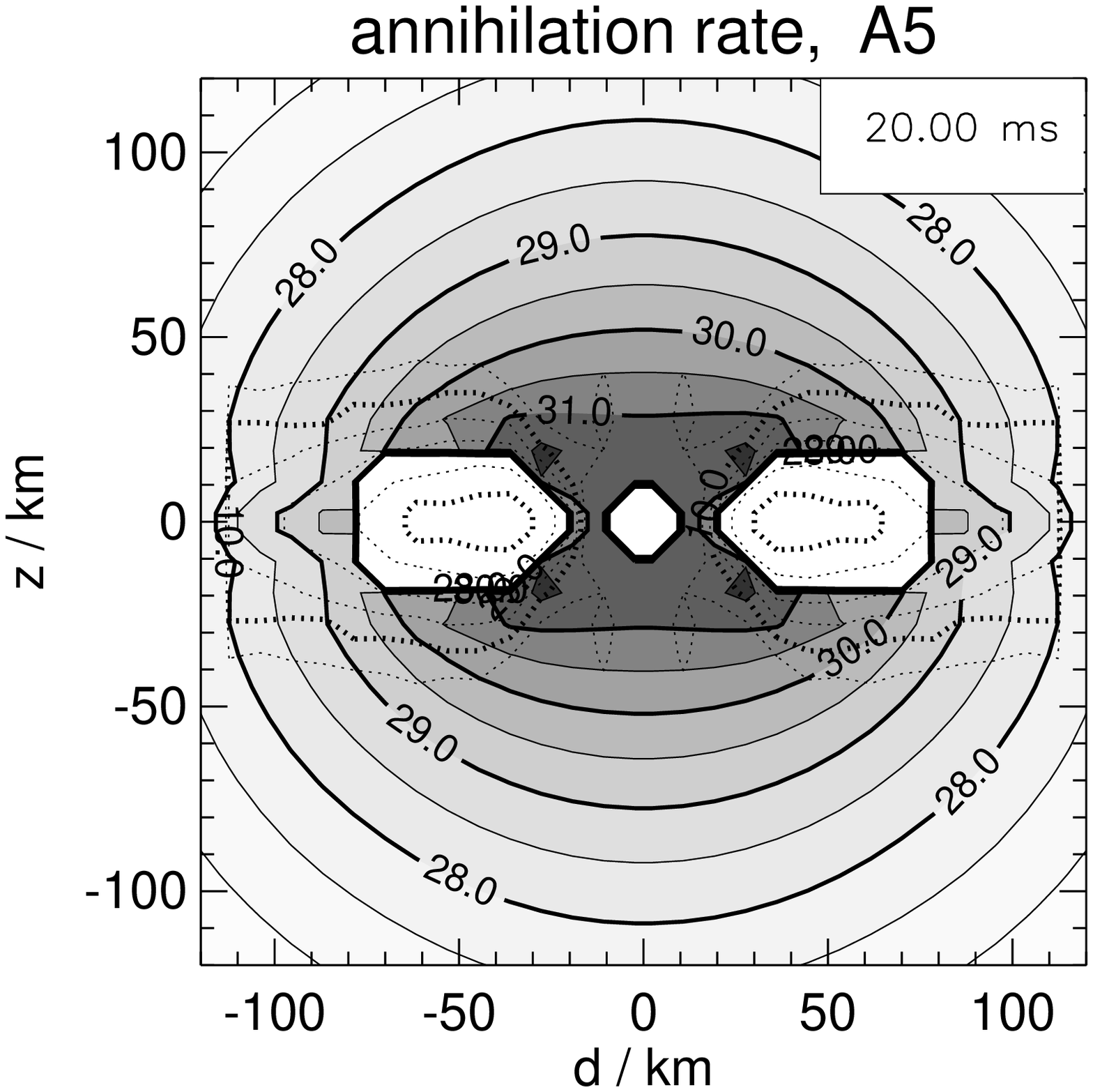,width=0.49\textwidth}}
}
\caption[]{\small 
Azimuthally averaged maps of the rate density
of energy deposition by $\nu\bar\nu$ annihilation into $e^+e^-$ pairs
(in ${\rm erg\,cm}^{-3}{\rm s}^{-1}$, contours represent logarithms,
the levels are spaced in steps of 0.5~dex)
in the surroundings of the accreting BH (white octagonal area 
at the center) which originates from the merging of two NS's with 
$1.6\,M_{\odot}$ baryonic mass (left) or of a $1.6\,M_{\odot}$ NS
with a $5\,M_{\odot}$ BH (right). In the former case the accretion
disk (AD) has a mass of $0.26\,M_{\odot}$ and the total energy deposition
rate below a density of $10^{11}\,{\rm g\,cm}^{-3}$ is 
$4.9\times 10^{50}\,{\rm erg\,s}^{-1}$. In the latter case the
AD mass is $0.45\,M_{\odot}$ and the integral energy deposition rate
is 15 times higher, $71.6\times 10^{50}\,{\rm erg\,s}^{-1}$. The dotted
lines represent isodensity contours of the AD, the dashed lines show
the neutrinospheres for the different $\nu$ and $\bar\nu$
flavors. The time after the start of the
simulation is given in the upper right corner.
}
\label{janka.fig3}
\end{figure}

Our simulations
confirm suggestions that these events could be the sources of a
subclass of short gamma-ray bursts at cosmological 
distances (Fig.~\ref{janka.fig2}).
BH-NS mergers yield between 10 and 100 times more energy than
NS-NS mergers (Table~1). The annihilation
of neutrinos and antineutrinos (Fig.~\ref{janka.fig3})
which are emitted from the hot matter 
that is accreted into the (forming) black hole, yields total energies
$E_{\nu\bar\nu}$ up to $10^{51}\,{\rm erg}$ in our models.
It is therefore enough
efficient to explain observable burst luminosities 
$L_{\gamma}\sim E_{\nu\bar\nu}/(f_{\Omega}t_{\gamma})$ up to 
several $10^{53}\,{\rm erg\,s}^{-1}$ for burst durations 
$t_{\gamma}\approx 0.1$--1$\,$s, if the $\gamma$ emission is 
beamed in two moderately focussed jets into a fraction
$f_{\Omega} = 2\delta\Omega/(4\pi)\approx 1/100$---$1/10$ of the sky.
Such a modest amount of beaming (jet opening half-angles between about
ten and several ten degrees) has to be expected because the
dense gas in the equatorial plane of the accreting black hole 
prevents relativistic expansion of the pair-plasma jets outside a 
low-density funnel along the system axis. More energy for the burst
can be available if Kerr effects of the rotating black hole are taken
into account.

The evolution tracks of compact binaries shown in Fig.~\ref{janka.fig2} 
depend on the NS and BH masses, on
the total angular momentum in the system, and on the stiffness of the
nuclear equation of state. Energetic relativistic pair-plasma
jets can be produced when a hot accretion disk forms around
the BH, because high neutrino fluxes are emitted over a time much
longer than the dynamical timescale of the system. This allows for
high efficiencies $q_{\nu}$ and $q_{\nu\bar\nu}$ and a large total
energy conversion by $\nu\bar\nu\to e^+e^-$.
Gamma-ray bursts from these scenarios are accompanied by the
ejection of nucleosynthesis products in considerable amounts,
the emission of large numbers of neutrinos with MeV energies,
and the production of strong gravitational-wave signals.

%
%
%
%

\acknowledgements{MR is grateful for support by a PPARC Advanced Fellowship, 
HTJ acknowledges support by the DFG on grant 
``SFB 375 f\"ur Astro-Teilchenphysik''.}

\vfill
\end{document}